\def\presentation{
\voffset -.50in  \hoffset -.19in
\oddsidemargin 0in \evensidemargin 0in
\marginparwidth .75in \marginparsep 7pt \topmargin 0in
\headheight 12pt \headsep .25in
\footheight 18pt \footskip .35in
\textheight 9.5in \textwidth 6.5in
\columnsep 10pt \columnseprule 0pt }
\documentstyle[10pt]{article}
\presentation
\begin{document}
%

%
\def\tilde{\widetilde}
\def\bar{\overline}
\def\hat{\widehat}
\def\*{\star}
\def\[{\left[}
\def\]{\right]}
\def\({\left(}
\def\){\right)}
\def\zb{{\bar{z} }}
\def\frac#1#2{{#1 \over #2}}
\def\inv#1{{1 \over #1}}
\def\half{{1 \over 2}}
\def\d{\partial}
\def\der#1{{\partial \over \partial #1}}
\def\dd#1#2{{\partial #1 \over \partial #2}}
\def\vev#1{\langle #1 \rangle}
\def\bra#1{{\langle #1 |  }}
\def\ket#1{ | #1 \rangle}
\def\rvac{\hbox{$\vert 0\rangle$}}
\def\lvac{\hbox{$\langle 0 \vert $}}
\def\2pi{\hbox{$2\pi i$}}
\def\e#1{{\rm e}^{^{\textstyle #1}}}
\def\grad#1{\,\nabla\!_{{#1}}\,}
\def\dsl{\raise.15ex\hbox{/}\kern-.57em\partial}
\def\Dsl{\,\raise.15ex\hbox{/}\mkern-.13.5mu D}
\def\comm#1#2{ \BBL\ #1\ ,\ #2 \BBR }
\def\x{\stackrel{\otimes}{,}}
\def\det{ {\rm det}}
\def\tr{{\rm tr}}
%
%
\def\th{\theta}         \def\Th{\Theta}
\def\ga{\gamma}         \def\Ga{\Gamma}
\def\be{\beta}
\def\al{\alpha}
\def\ep{\epsilon}
\def\la{\lambda}        \def\La{\Lambda}
\def\de{\delta}         \def\De{\Delta}
\def\om{\omega}         \def\Om{\Omega}
\def\sig{\sigma}        \def\Sig{\Sigma}
\def\vphi{\varphi}      \def\ee{{\rm e}}
%
%
\def\CA{{\cal A}}       \def\CB{{\cal B}}       \def\CC{{\cal C}}
\def\CD{{\cal D}}       \def\CE{{\cal E}}       \def\CF{{\cal F}}
\def\CG{{\cal G}}       \def\CH{{\cal H}}       \def\CI{{\cal J}}
\def\CJ{{\cal J}}       \def\CK{{\cal K}}       \def\CL{{\cal L}}
\def\CM{{\cal M}}       \def\CN{{\cal N}}       \def\CO{{\cal O}}
\def\CP{{\cal P}}       \def\CQ{{\cal Q}}       \def\CR{{\cal R}}
\def\CS{{\cal S}}       \def\CT{{\cal T}}       \def\CU{{\cal U}}
\def\CV{{\cal V}}       \def\CW{{\cal W}}       \def\CX{{\cal X}}
\def\CY{{\cal Y}}       \def\CZ{{\cal Z}}
%
%
\font\numbers=cmss12
\font\upright=cmu10 scaled\magstep1
\def\stroke{\vrule height8pt width0.4pt depth-0.1pt}
\def\topfleck{\vrule height8pt width0.5pt depth-5.9pt}
\def\botfleck{\vrule height2pt width0.5pt depth0.1pt}
\def\Zmath{\vcenter{\hbox{\numbers\rlap{\rlap{Z}\kern
		0.8pt\topfleck}\kern
		2.2pt \rlap Z\kern 6pt\botfleck\kern 1pt}}}
\def\Qmath{\vcenter{\hbox{\upright\rlap{\rlap{Q}\kern
		   3.8pt\stroke}\phantom{Q}}}}
\def\Nmath{\vcenter{\hbox{\upright\rlap{I}\kern 1.7pt N}}}
\def\Cmath{\vcenter{\hbox{\upright\rlap{\rlap{C}\kern
		   3.8pt\stroke}\phantom{C}}}}
\def\Rmath{\vcenter{\hbox{\upright\rlap{I}\kern 1.7pt R}}}
\def\Z{\ifmmode\Zmath\else$\Zmath$\fi}
\def\Q{\ifmmode\Qmath\else$\Qmath$\fi}
\def\N{\ifmmode\Nmath\else$\Nmath$\fi}
\def\C{\ifmmode\Cmath\else$\Cmath$\fi}
\def\R{\ifmmode\Rmath\else$\Rmath$\fi}
\def\cadremath#1{\vbox{\hrule\hbox{\vrule\kern8pt\vbox{\kern8pt
			\hbox{$\displaystyle #1$}\kern8pt} 
			\kern8pt\vrule}\hrule}}
\def\proof{\noindent {\underline {Proof}.} }
\def\cqfd{ {\hfill{$\Box$}} }
\def\square{ {\hfill \vrule height6pt width6pt depth1pt} } 
%
%
\def\debut{ \begin{eqnarray} }
\def\fin{ \end{eqnarray} }
\def\non{ \nonumber }
%

%
%
\rightline{SPhT/98/036, IHES/P/98/31}
\vskip 2cm
\centerline{\LARGE\bf Scaling and Exotic Regimes}
\vskip 0.2cm
\centerline{\LARGE\bf in Decaying Burgers Turbulence}
\vskip 1.5cm
\centerline{\large  Denis Bernard${}^{a}$\footnote[1]{Membre 
du CNRS; dbernard@spht.saclay.cea.fr} 
and Krzysztof Gaw\c{e}dzki${}^{b} $\footnote[2]{Membre 
du CNRS; bongo@ihes.fr}    }
\bigskip 
\centerline{${}^{a}$ Service de Physique Th\'eorique 
de Saclay\footnote[3]{Laboratoire de la Direction des Sciences 
de la Mati\`ere du Commisariat \`a l'Energie Atomique.}}
\centerline{F-91191, Gif-sur-Yvette, France.}
\medskip
\centerline{${}^b$ Institut des Hautes Etudes Scientifiques}
\centerline{F-91440, Bures-sur-Yvette, France.}
\medskip
\vskip2.5cm
\centerline{\bf{Abstract}}
\

We analyze the stochastic scaling laws arising in the invicid
limit of the decaying solutions of the Burgers equation.
The linear scaling of the velocity structure functions is shown 
to reflect the domination by shocks of the long-time asymptotics. 
We exhibit new self-similar statistics of solutions describing 
phases with diluted shocks. Some speculations are included on the nature 
of systems whose large time behavior is described by the new 
statistics.

\vfill
\newpage
%
%
\section{Introduction}
The Burgers equation, a version of the Navier-Stokes equation
without pressure, takes in the 1+1 dimensions the form:
\debut
\d_t u + u \d_x u - \nu \d_x^2 u = 0 \label{burgers}
\fin
where $u= u(x,t)$ is the velocity field. We have not
included the force term since we shall be interested
in the free decay of initial data. Although we shall stick
to the 1-dimensional space, most of the following can be generalized 
to higher dimensions. We are interested in statistical 
properties of the velocity field at time $t>0$, given the statistics 
of random initial data. The equation (\ref{burgers}) has
the $u(t,x)\mapsto-u(t,-x)$ symmetry which, if respected
by the statistics of the initial conditions, will persist
at all times. The problem is to evaluate the $n$-point correlation 
functions $\vev{\prod_j u(x_j,t)}$ at equal time in the invicid limit 
$\nu\to 0$. At large time these correlation
functions are expected to flow towards some `universal' functions 
manifesting a self-similar character. In other words:
\debut
\vev{\prod_{j=1}^n u(x_j,t)} ~\simeq ~ u^n(t)~ B_n\({\frac{x_j}{l(t)}}\)
\quad~~~~~~~~~~ {\rm for}\quad t\quad {\rm large} \label{asymp}
\fin
with $l(t)$ and $u(t)\simeq \d_t l(t)$ being the  characteristic length 
and the characteristic velocity at time $t$.
Note the order of the limits: first $\lim\limits_{\nu \to 0}$ and
then $\lim\limits_{t \to \infty}$. These limits do not commute.
Each asymptotic universal statistics, which are specified
by their correlation functions $B_n(x_j/l(t))$, have their own basin 
of attraction.  
\vskip 0.2cm

If the initial statistics are Gaussian with zero mean,
they are encoded in the initial velocity two-point 
function $\Ga(x-y)=\vev{u_0(x)u_0(y)}$ which we assume translation 
invariant. As is known since the work of Burgers, the large time 
behavior depends crucially on whether $\CJ= \int dx\, \Ga(x)$ vanishes 
or not. The case $\CJ\not= 0$ was the case first studied by Burgers 
himself \cite{burg}. In that case the large time behavior is governed 
by a self-similar solution with characteristic length $l(t)\sim 
t^{2/3}$. The case $\CJ=0$ was analyzed by Kida in his important 
paper \cite{kida}. It is convenient to introduce the potential 
$\Phi(x,t)$ such that $u(x,t)=\d_x\Phi(x,t)$. If $\CJ=0$, we may 
assume that the initial potential is Gaussian with mean zero
and the two-point function 
\debut
G(x-y) = \vev{\Phi_0(x)\Phi_0(y)}\,.        \label{defG}
\fin
Of course $\Ga(x)=-\d_x^2 G(x)$. Assuming {\it a priori} that 
the minima of the initial potential are independent, 
Kida showed in  ref.\,\cite{kida} that the large time behavior has 
a characteristic length $l(t)\sim t^{1/2}$ up to a logarithmic
correction.  A more precise formulation of this statement, proved in
ref.\cite{Moetal} under very mild hypothesis on $G(x)$, 
(e.g. for $G(x)$ being a smooth function decreasing rapidly 
enough at infinity, c.f. also \cite{Aur} and references therein)
states that the following limit exists:
\debut
\lim_{\ep\to 0} |\log\ep|^\half~
u\({\frac{x}{\ep},\frac{t}{\ep^2}|\log\ep|^\half }\)
\ \cong\ u_K(x,t)\,. \label{limkida}
\fin
Here and in the following, $\cong$ means an equality in law,
i.e. inside any correlation functions. 
The statistics of the limiting velocity field $u_K(x,t)$
was explicitly constructed in \cite{kida,Moetal}.
$u_K(x,t)$ is self-similar with a diffusive scaling, 
$l_K(t)= t^{1/2}~\De^{1/4}$ where $\Delta=G(0)$. 
Note that one of the hypothesis for having Kida's statistics 
at large time is that $0<\De<\infty$, i.e. that the initial potential 
two-point function is regular at the origin.
\vskip 0.2cm

In this letter, we shall construct other self-similar solutions of
decaying Burgers turbulence. Although these solutions are different
from Kida's statistics, they share in commun the fact to be constructed
from Poisson point processes. These solutions may be relevant 
in the large time behavior of systems whose initial correlation 
functions are singular at coinciding points.
\vskip 0.2cm

In Sect.\,2, we introduce a few basic facts concerning the Burgers 
equation and we describe some universal features of fields localized 
on shocks. In Sect.\,3, we construct the self-similar statistics 
and we prove that they effectively are solutions of the turbulent 
problem. We also give a few more detailed informations on a particular 
case. Comments and speculations are gathered in Sect.\,4.

{\bf Acknowledgements}. We would like to thank Antti Kupiainen
for the collaboration in initial stages of this work 
and to Marc M\'{e}zard for stimulating discussions.

\section{Basic facts about the Burgers equation}
In order to fix notations, we recall few elementary facts concerning 
the Burgers equation \cite{burg,kida}.
As is well known, the equation is solved by implementing the Cole-Hopf 
transformation which maps it to the heat equation. 
This works as follows.
Let $Z(x,t)=\exp[-\inv{2\nu}\Phi(x,t)]$ where $u(x,t)=\d_x \Phi(x,t)$.
Eq.\,(\ref{burgers}) for $u$ is mapped into the heat equation for $Z$:
\debut
\[{\d_t - \nu \d_x^2}\] Z(x,t)=0\,. \non
\fin
Thus, given the initial condition $u(x,t=0)\equiv u_0(x)$, the velocity
field at a later time $t$ is recovered from the potential $\Phi(x,t)$
given by the relation
\debut
\exp\[{-\inv{2\nu}\Phi(x,t)}\] = \int \frac{dy}{\sqrt{4\pi \nu t}}
\exp\[{ -\inv{2\nu}\({\Phi_0(y) +\frac{(x-y)^2}{2t} }\) }\] \label{solu}
\fin
with $\Phi_0(x)$ standing for the initial potential such that 
$u_0(x)=\d_x\Phi_0(x)$. The invicid Burgers equation  corresponds 
to the limit $\nu \to 0$. The solution is then given by solving 
a minimalization problem: 
\debut
u(x,t) = \d_x\Phi(x,t) \quad {\rm with}\quad
\Phi(x,t)= \min_y \({\Phi_0(y) +\frac{(x-y)^2}{2t} }\). \label{minsol}
\fin
Outside shocks the minimum is reached for only one value $y_*$ of $y$,
the solution of the equation $u_0(y_*)t=x-y_*$. The velocity is 
$u(x,t)=\frac{x-y_*}{t}=u_0(y_*)$. It is effectively a local solution
of the invicid Burgers equation since, by the minimum condition
defining $y_*$, we have $u(x,t)= u_0(x- t u(x,t))$. 
A simple geometrical construction of the solution (\ref{minsol}) 
is described in refs.\cite{burg,kida}. For large $t$, $y_*$ coincides 
approximately with one of the local minima of $\Phi_0(y)$ and
it practically does not change under small variations of $x$
so that, in between the shocks, the velocity is approximately 
linear with the slope $\inv{t}$.
\vskip 0.2cm

Shocks appear when the minimum is reached for two values $y_1$ and 
$y_2$ of $y$. Let $\Phi_{1,2}=\Phi_0(y_{1,2})$ be the value 
of the initial potential at these points. Then eq.\,(\ref{solu}) 
allows one to determine the velocity profil $u_{s}(x,t)$ around 
and inside the shocks at finite value of the viscosity $\nu$ 
by expressing $\exp\[{-\inv{2\nu} \Phi_{s}(x,t)}\]$ as the sum 
of contributions from the two minima. One obtains:
\debut
u_{s}(x,t) = \inv{t}\({x-\half(y_1+y_2)}\) - \frac{\mu_{s}}{2t} \ 
\tanh\({ \frac{\mu_{s}}{4\nu t}\({x-\xi_{s} t -\half(y_1+y_2)}\)}\) 
\label{ushock}
\fin
where $\mu_{s}=y_1-y_2>0$ and $\xi_{s}= \frac{\Phi_1-\Phi_2}{y_1-y_2}$.
In the invicid limit $\nu\to 0$, this becomes
\debut
u_{s}(x,t)|_{_{\nu=0}} =\, \xi_s+\frac{x-x_s(t)}{t}
-\frac{\mu_s}{2t}\bigg(\theta(x-x_s(t))-\theta(x_s(t)-x)\bigg)
\label{ushock0}
\fin
where $x_{s}(t)=\xi_{s} t +\half(y_1+y_2)$ is the time $t$ position
of the shock which moves with the velocity $\xi_{s}$ and follows 
a Lagrangian trajectory. $\th(x)$ is the step function.
The values of the velocity on the two sides of the shock are:
\debut
u_{s}^\pm = u_s(x_{s}\pm 0)=\xi_{s} \mp \frac{\mu_{s}}{2t} 
\label{shock+-}
\fin
so that $\mu_{s}\over t$ is the amplitude of the shock.
\vskip 0.2cm

The presence of shocks is at the origin of universal features 
which are independent of the details of the statistics. They may be 
analyzed by looking at fields localized on the shocks. By definition, 
these fields may be represented for any realization as:
\debut
\CO_g (x,t) = \sum_{shocks}~ g(\xi_{s},\mu_{s})~
\de(x-x_{s}(t)) \label{loc}
\fin
where the sum is over the shocks with $x_{s}(t)$ denoting the position
of the shock, $\xi_{s}$ its velocity and $\mu_{s}\over t$ its amplitude.
These fields are labeled by functions of $\xi_{s}$ and $\mu_{s}$.
By using the velocity profile (\ref{ushock}) inside and around 
the shocks, we may map fields defined in terms of the velocity 
$u(x,t)$ into the shock representation.
For example, the shifted derivative of the velocity field 
$\(\d_x u(x,t)-\inv{t}\)$ is for large $t$ localized on the shocks
since away from shocks, $u(x,t)=\frac{x-y_*}{t}$ with $y_*$ 
almost independent of $x$. More generally, the generating 
function $\({\d_x-\frac{\la}{t}}\) \exp\[{\la\, u(x,t)}\]$ also becomes
localized on the shocks for large $t$. Using the velocity profiles 
(\ref{ushock}) or directly (\ref{ushock0}), one finds its shock 
representation:
\debut
\({\d_x-\frac{\la}{t}}\) \ee^{\la\, u(x,t)}= -2\sum_s \ee^{\la\xi_{s}}
\sinh(\frac{\la\mu_{s}}{2t})~\de(x-x_{s}(t))\,. \label{dula}
\fin
Note that this differs from the  products 
$\la(\d_x u(x,t)-\frac{1}{t})\, \ee^{\la\, u(x\pm 0,t)}$ which are evaluated
using the fact that the velocity on the two sides of a shock
are $u_s^\pm = \xi_{s} \mp \frac{\mu_{s}}{2t}$: 
\debut
\la\({\d_xu(x,t)-\frac{1}{t}}\) \ee^{\la\, u(x\pm 0,t)}=
-\la\sum_s \frac{\mu_{s}}{t} \exp\[{\la(\xi_{s} \mp \frac{\mu_{s}}{2t})}\]
\de(x-x_{s}(t))\,.
\label{u+-}
\fin
\vskip 0.2cm

Another example of a field localized on shocks is provided by 
the dissipation field $\ep(x,t)$ defined by
$\ep(x,t) = \lim\limits_{\nu \to 0} \nu (\d_xu)^2$.
As is well known, $\ep(x)$, which is naively zero due to
the prefactor $\nu$ in its definition, is actually a non-trivial
field since $(\d_x u)^2$ is singular in the invicid limit. 
Integrating $\nu (\d_xu_s)^2$ around the shock one obtains in the
limit $\nu\to0$ the contribution $\inv{12}({\mu_{s}\over t})^3$. 
Hence the shock representation of $\ep(x)$ is:
$$\ep(x)= \inv{12} \sum_s (\frac{\mu_s}{t})^3\, \de(x-x_{s}(t))\,.$$
More generally one finds, using again the velocity profile 
(\ref{ushock}), the shock representation of the generating
function $\ep(x,t)\, \ee^{\la\, u(x,t)}$. Namely:
\debut
\ep_\la(x,t)\equiv\ep(x,t)\, \ee^{\la\, u(x,t)}
\,=\, 2\lambda^{-3}\sum_{c} \ee^{\la\xi_{s}}\({\frac{\la\mu_s}{2t}
\cosh(\frac{\la\mu_s}{2t}) - \sinh(\frac{\la\mu_s}{2t}) }\)
\de(x-x_{s}(t))\,. \label{epla}
\fin
Note that for $\nu\not=0$, the Burgers equation (\ref{burgers}) 
implies that
\debut
0\,=\,\Big(\d_t\,+\,\lambda\,\d_\lambda\,\inv{\la}\,\d_x
-\la\,\nu\,(\d_x^2u)\Big)\, \ee^{\lambda\, u}\,=\,
\Big(\d_t\,+\,\lambda\,\d_\lambda\,\inv{\la}\,\d_x
+\la^2\,\nu\,(\d_xu)^2-\nu\, \d_x^2\Big)\, \ee^{\lambda\, u}\,.
\fin
Since $\d_x^2 \ee^{\la\, u}$ has a (distributional)
limit when $\nu\to 0$, we may expect that at $\nu=0$ 
\debut
\Big(\d_t\,+\,\lambda\,\d_\lambda\,\inv{\la}\,\d_x\Big)\, 
\ee^{\lambda\, u}\,+\,\lambda^2\,\epsilon_\lambda\,=\, 0
\label{invbur}
\fin
encoding the invicid version of the Burgers equation.
Indeed, eq.\,(\ref{invbur}) may be verified directly
at $\nu=0$ by computing $\Big(\d_t\,+\,\lambda\,\d_\lambda\,
\inv{\la}\,\d_x\Big)\, \ee^{\lambda\, u}$ with the use 
of the limiting profile (\ref{ushock0}). 
\vskip 0.2cm

Comparison of eq.\,(\ref{epla}) with eqs.\,(\ref{dula},\ref{u+-})
yields an alternative representation of the dissipation field
in terms of the velocity field:
\debut
\ep_\la(x,t)&=&\inv{2}\lambda^{-2}\,(\d_x u(x,t))
\({2 \ee^{\la\, u(x,t)} - \ee^{\la\, u(x+0,t)}- \ee^{\la\, u(x-0,t)} }\)
\label{eprep}\\
&=& \inv{2}\lambda^{-2}\,(\ee^{\la\, u(x,t)})
\Bigl({2\d_x u(x,t)-\d_xu(x+0,t)-\d_x u(x-0,t)}\Bigr) \non
\fin 
Eq.\,(\ref{eprep}) is an extension of the well-known formula
$\ep(x)= \inv{12} \lim\limits_{l\to 0}\d_l\[{u(x)-u(x+l)}\]^3$.
As expected and manifest in eq.\,(\ref{eprep}), the dissipation 
field is located on the discontinuity of the derivative 
of the velocity field. Eq.\,(\ref{eprep}) does not coincide 
with the operator product expansion suggested in \cite{polya} 
for the forced Burgers turbulence and expressing $\ep_\la$ as 
a combination of $\ee^{\la\, u}$ and $\d_x \ee^{\la\, u}$.
\vskip 0.2cm

Fields localized on shocks form a closed algebra.
When shocks are diluted, these operators are expected
to satisfy a simple operator product expansion:
\debut
\CO_f(x,t)\cdot \CO_g(y,t) = \de(x-y)~ \CO_{fg}(x,t)\ 
+\ {\rm regular}\,. \non
\fin
The contact term $\de(x-y)$ in this operator product expansion arises
from the coinciding shocks in the double sum representing the product 
operator. As an application, let us present an argument showing that 
the structure functions scale linearly in $x$. Indeed, using 
the representation (\ref{dula}) for the operator $\({\d_x-\la}\) 
\ee^{\la\, u(x,t)}$ at $t=1$, one finds:
\debut
\({\d_x}-\la\)^2\vev{ \ee^{\la\, (u(x)-u(0))}} = \de(x)~\vev{\CO_\varphi}
\ +\  {\rm regular} \non
\fin
with $\CO_\varphi(x)= 2\sum_s (\cosh(\la\mu_s)-1)\,\de(x-x_{s})$.
By integrating, this implies:
\debut
\vev{ \ee^{\la (u(x)-u(0))}} = 1 + a(\la)~ x + 
\frac{\vev{\CO_\varphi}}{2}~|x| \ +\  o(|x|)
\label{sd}
\fin
with $a(\la)=-a(-\la)$. Eq.\,(\ref{sd}) implies that, at short distances,
$\vev{(u(x)-u(0))^n}$ is proportional to $|x|$ for even positive $n$: 
\debut
\vev{(u(x)-u(0))^n}= \vev{\CO_{\mu^n}}~|x| ~~+\  o(|x|)\non
\fin
and it is consistent with the behavior
proportional to $x$ for odd $n>1$, as the one holding for the 3-point 
function. In words, the anomalous scalings of the structure functions 
in Burgers turbulence are a simple echo of the shocks. They are universal 
(at least when shocks are diluted): only the amplitudes are statistics 
dependent. Some of these representations and formal manipulations also 
apply to the forced Burgers equation.

\section{Self-similar solutions}
Self-similar behavior such as in eq.\,(\ref{asymp}) will be true 
at any time, i.e. not only asymptotically, if the initial correlation 
functions scale. Indeed, from the explicit solution (\ref{minsol}) 
it immediately follows that (for the Gaussian initial data)
\debut
G_0(s x) = s^{2h}\, G_0(x)\quad \Longrightarrow \quad 
s^{1-h}\, u(sx,\, s^{2-h} t)\ \cong\ u(x,t)\,. \label{scal}
\fin
$h$ is the dimension of the initial potential.  Such scaling 
behavior corresponds to a characteristic length $\l(t)\sim 
t^{\inv{2-h}}$.
For this length to grow with time we must have $h<2$.  Shocks are 
expected to be dense for $1<h<2$ and diluted for $h<1$ 
\cite{sinai,she}.
\vskip 0.2cm

Demanding a self-similar behavior for the correlation functions
imposes constraints on the correlation functions of the
velocity field:
\debut
\Big[(2-h)\, t\d_t+\sum_j\(x_j\d_{x_j}+(1-h)\la_j\d_{\la_j}\)\Big] 
\vev{\prod_k \ee^{\la_k u(x_k,t)}}\ =\ 0\,.
\label{sca}
\fin

One may construct self-similar solutions with scaling
dimensions $h$ for $0\geq h>-1$ by generalizing the 
representation of Kida's asymptotic solution \cite{kida}.
By construction, the velocity $u_h(x,t)$ has the following
form:
\debut
u_h(x,t)=\d_x \Phi_h(x,t) \quad {\rm with}\quad
\Phi_h(x,t) = \min_j\({ \phi_j + \frac{(x-y_j)^2}{2t} }\) 
\label{upoisson}
\fin
where $(\phi_j,y_j)_{j\in {\bf Z}}$ is a Poisson point process 
with intensity $f_h(\phi)d\phi dy$. 
Recall that this means that the probability to find
a point of this process in an infinitesimal cell centered
at $(\phi,y)$ is $f_h(\phi,y)\, d\phi dy$ and that such elementary 
events are independent. To assure the translation invariance, 
$f_h$ will depend only on $\phi$. For any given realization, 
the velocity field (\ref{upoisson}) has an exact sawtooth profile
with slope $\inv{t}$. In this Ansatz all shocks are created 
at time $t=0$. The later time evolution is then governed 
by the shock collisions: the biggests eating the smallests.
\vskip 0.2cm

Let us first show that eq.\,(\ref{upoisson}) is preserved 
by the evolution specified by the invicid Burgers equation.
At a time $t'=t+\tau>t$, the velocity field $u_h(x,t')$ is
given by:
\debut
u_h(x,t+\tau) = \d_x \min_y \({ \Phi_h(y,t) 
+ \frac{(x-y)^2}{2\tau} }\). \non
\fin
Inserting the expression (\ref{upoisson}) for $\Phi(y,t)$ and
commuting the minimization over $y$ and over $j$'s, we get:
\debut
u_h(x,t+\tau) &=& \d_x \min_j \({ \phi_j + 
\min_y\({ \frac{(y-y_j)^2}{2t}+ \frac{(x-y)^2}{2\tau} }\) }\) \non\\
&=& \d_x \min_j\({ \phi_j + \frac{(x-y_j)^2}{2(t+\tau)} }\). \non
\fin
\vskip 0.2cm

Next we determine the intensity $f_h(\phi)\, d\phi dy$
such that the solution (\ref{upoisson}) is self-similar
with scaling dimension $h$, i.e.
$s^{1-h}\, u_h(s x,\,s^{2-h}t)\, \cong\, u(x,t)$.
Let us spell out this condition for the potential $\Phi_h(x,t)$.
By definition,
\debut
s^{-h} \Phi_h(s x,\, s^{2-h}t) &=&
\min_j \({ s^{-h} \phi_j + \frac{(x-s^{-1} y_j)^2}{2t} }\) \non\\
&=& \min_j \({ \hat \phi_j + \frac{(x-\hat y_j)^2}{2t} }\) \non
\fin
where $\hat \phi_j=s^{-h} \phi_j$ and $\hat y_j = s^{-1} y_j$.
Since $u_h(x,t)=\d_x \Phi_h(x,t)$,
demanding self-similarity amounts to require
that $s^{-h} \Phi_h(s x,\, s^{2-h}t)\cong \Phi_h(x,t)-C_s\,$
with $C_s$ a constant. This equality will be true in law if the
intensity is such that:
$f_h(\phi)\, d\phi dy = f_h(\hat \phi + C_s)\, d\hat \phi d\hat y$.
Up to an irrelevent translation of $\phi$ the solutions
of this equation are:
\debut
f_h(\phi) &=&\hbox to 5.5cm{$\cases{\rm{const.}\,
\phi^{-(\frac{1+h}{h})}& for $\phi\geq 0$\cr
		      0    & for $\phi \leq 0$\cr}$\hfill}
{\rm with}\quad -1<h<0\,, \label{intens1}\\
f_0(\phi) &=&\hbox to 5.5cm{$\exp[{\rm const.}\,\phi]$\hfill}
{\rm with}\quad h=0\,. \label{intens2} 
\fin
Thus we showed that the representation (\ref{upoisson})
of the velocity field in terms of the Poisson process 
$(\phi_j,y_j)$ with intensity (\ref{intens1}) is (i) self-similar 
and (ii) preserved by the evolution. In particular the relations 
such as eqs.\,(\ref{sca}) will be satisfied. Moreover
the invicid form (\ref{invbur}) of the Burgers equation
for each realization implies the Hopf equations 
for the correlators:
\debut
\Big[\d_t+\sum\limits_j\la_j\, \d_{\la_j}\,\inv{\la_j}\, \d_{x_j}\Big]
\vev{\prod\limits_k \ee^{\la_k u(x_k,t)}}
\,+\,\sum\limits_j\la_j^2\,\vev{\ep_{\lambda_j}\prod\limits_{k\not=j} 
\ee^{\la_k u(x_k,t)}}\ =\ 0
\label{hopf}
\fin
The time derivative may be eliminated from both equations leading
to the fixed-time version of the Hopf equations. The case $h=0$ 
corresponds to Kida's asymptotic solution. As pointed out in \cite{mez} 
it generalizes the Gumbel class of extreme statistics. The case 
$h\not=0$ should, correspondingly, generalize the Weibull class 
of extreme statistics.
\vskip 0.2cm

Let us describe in more details the case $h=-\half$.
It corresponds to initial potential correlation functions 
homogeneous of degree $-1$, e.g. like the Dirac delta function.
We denote by $\Phi^*(x,t)$ and $u^*(x,t)$ the corresponding
potential and velocity field. The statistics of the Poisson point
process $(\phi_j,y_j)$ in the representation (\ref{upoisson})
is specified by the intensity 
\debut
f^*(\phi)\ =\,\cases{ D^{-1}~\phi\, d\phi dy& for $\phi\geq 0$\,,\cr
		      0    & for $\phi \leq 0$\,.}
\label{statint}
\fin
$D$ is a constant with dimension $(length)^5\times (time)^{-2}$.
Recall that the velocity field $u^*(x,t)$ 
is such that $u^*(x,t)\,\cong\, t^{-3/5} u^*(xt^{-2/5},1)$.
In other words, the characteristic length at time $t$ is
$l^*(t)= D^{1/5}~ t^{2/5}$.
\vskip 0.2cm

The one-point function of the potential scales
as $t^{-2/5}$ and thus diverges at $t=0$ (it does not contribute
to the one point function of the velocity which vanishes). 
The two-point function $G^*(x,t)$ of $\Phi^*(x,t)$ satisfies 
$G^*(x,t)=t^{-2/5}G^*(xt^{-2/5},1)$. Since, as we shall see, 
$G^*(x,1)$ is smooth, regular at the origin and decreasing
exponentially at infinity, at zero time $G(x,t)$
becomes proportional to the Dirac delta function, 
\debut
\lim_{t\to 0} G^*(x,t) = {\rm const.} ~\de(x)\,. \non
\fin
\vskip 0.2cm

Once the velocity field has been parametrized in terms 
of Poisson processes as in eq.\,(\ref{upoisson}), 
it is easy to compute any correlation 
functions. For example, the one-point generating function is
\debut
\vev{\ee^{\la\, u^*(x,t)}} = \int d\phi dy\, \CP_x(\phi,y)~ 
\ee^{\la(x-y)/t} \non
\fin
with $\CP_x(\phi,y)$ denoting the probability (density) 
of $(\phi,y)=(\phi_{j^*},y_{j^*})$ for $j^*$
minimizing $\({\phi_j+ \frac{(x-y_j)^2}{2t}}\)$. 
To compute this probability, imagine dividing 
the $(\phi,y)$-plane into elementary cells of size $d\phi dy$ 
with the probability for a point $(\phi_j,y_j)$ to be 
in the cell centered at $(\phi,y)$ equal to $f(\phi) d\phi dy$. 
Thus, the probability $\CP_x(\phi,y)$ is equal to the product 
of the probability for the elementary cell centered around 
$(\phi,y)$ to be occupied by a point of the process 
times the probability for the other cells around $(\phi',y')$ 
s.t. $\phi'+\frac{(x-y')^2}{2t}<\phi+\frac{(x-y)^2}{2t}$
to be empty (if this condition is violated, the cell 
around $(\phi',y')$ may be either occupied or empty). Hence
\debut
\CP_x(\phi,y)d\phi dy= f(\phi)d\phi dy 
\prod_{d\phi' dy'}\Bigl(1\,-\,\chi(\phi',y';\phi,y) f(\phi')
d\phi' dy' \Bigr) \non
\fin
where $\chi(\phi',y';\phi,y)$ is the characteristic
function of the constraint $\phi'+\frac{(x-y')^2}{2t}
<\phi+\frac{(x-y)^2}{2t}$.
Taking the continuum limit of infinitesimal cells and approximating
the product over the cells around $(\phi',y')$ by the exponential 
of a sum leads to the result:
\debut
\vev{\ee^{\la\, u^*(x,1)}} &=& \int d\phi dy~\ee^{-\la\, y}~ 
f(\phi-\frac{y^2}{2})\exp\[{-\int dz \int_{-\infty}^{\phi
-\frac{z^2}{2}} d\phi' f(\phi') }\] \non\\
&=& \frac{2}{|\la|^5} \int_0^{\infty} dX ~ X 
(X\cosh X-\sinh X)~\exp\[{-\frac{2}{15}(X/|\la|)^5}\]. 
\label{1point}
\fin
We have set $t=1$ and $D=1$. The dependence on these parameters 
is restored by replacing $\la$ by $\la D^{1/5} t^{-3/5}$. One
may check directly that the above expression for the 1-point 
function satisfies the identities (\ref{sca}) and (\ref{hopf}).
$\vev{\ee^{\la\, u^*(x,1)}}$ is regular around $\la=0$ and its 
behavior at infinity is:
\debut
\vev{\ee^{\la\, u^*(x,1)}} \simeq {\rm const}.~ |\la|^{-15/8}~
\exp[{- {\rm const}.~ |\la|^{5/4} }]
\quad {\rm for}\quad \la\to \infty\,.
\label{asymp1}
\fin
This has to be compared with Kida's statistics for which 
$\vev{\ee^{\la\, u^*(x,1)}}= \exp[-{\rm const}.~\la^2]$.
The two-point functions can be computed similarly. For $x>0$,
\debut
\vev{\ee^{\la(u^*(x)-u^*(-x))}} = \ee^{2\la x} 
\int_{\phi>0} d\phi dy \Bigl[{ \phi +
2x J_\la(\phi;y;x)J_\la(\phi;-y;x) }\Bigr]\ 
\ee^{-I_0(\phi;y;x)-I_0(\phi;-y;x)} \label{2point}
\fin
with
\debut
J_\la(\phi;y;x) &=&\int_\CD dz 
(\phi +\half(x+y)^2 -\half z^2)~\ee^{\la(z-x)}\,, \non\\
I_0(\phi;y;x) &=&\half\int_\CD dz 
(\phi +\half(x+y)^2 -\half z^2)^2 \non
\fin
where the domain of integration in both cases is $\CD
=\{\, z\ \vert\ z\leq y+x;\ z^2\leq 2\phi + (y+x)^2\}$.
Eq.\,(\ref{2point}) may be used to check that for $t\not= 0$
the two-point function $G(x,t)$ is smooth, fast decreasing 
at infinity and regular at the origin. There are no difficulties, 
but not much motivations, to compute in the same way the higher 
point correlation functions.

\section{Comments and speculations}
We have constructed self-similar solutions of the decaying
Burgers turbulence. It is natural to wonder if such
statistics effectively describe the long time behavior
of systems with smooth random initial data.
One may construct such examples
in a tautological way by taking as initial data
the potential $\Phi_0^\eta(x)$ obtained from the self-similar 
Ansatz (\ref{upoisson}) at small but non zero time:
$\Phi_0^\eta(x)=\Phi_h(x,\eta)$ with $\eta\not= 0$.
By construction, it defines a smooth initial statistics which
will have a large time asymptotics given by $\Phi_h(x,t)$.
Of course, this initial statistics is not Gaussian.
This shows, however, that the basin of attraction 
of the self-similar solution (\ref{upoisson}) is not totally 
empty.
\vskip 0.2cm

In the case $h=-\half$ the two-point function of $\Phi_0^\eta(x)$
tends to the Dirac delta function $\de(x-y)$ when $\eta\to 0$.
So the initial statistics $\Phi_0^\eta(x)$ may be thought 
of as a way to regularize an initial potential with $\de(x-y)$ 
two-point correlation function. Note that if we replace 
$\de(x-y)$ by a smooth cut-off dependent function 
$\De(x-y)$, the values at the origin $\De(0)$ diverge 
with the cut-off and Kida's asymptotic regime disappears
in the limit since its characteristics length diverges.
\vskip 0.2cm

The question is then whether there exist analogues
of eq.\,(\ref{limkida}) but with different respective scaling 
between $x$ and $t$ corresponding to non-zero values of $h$.
For example one may inquire about the existence of the limit
\debut
\Phi^{**}(x,t) \equiv \lim_{\ep \to 0}\,
\inv{\ep^{1/2}}\ \Phi\(\frac{x}{\ep},\frac{t}{\ep^{5/2}}\).  
\label{inter}
\fin
Naively, it corresponds to a limiting initial potential with
$\de(x-y)$ correlation function. Indeed, upon assuming that $\min_y$ 
and $\lim\limits_{\ep\to 0}$ commute, it follows from 
eq.\,(\ref{minsol}) that:
\debut
\lim_{\ep \to 0}~ \inv{\ep^{1/2}}\ \Phi\({\frac{x}{\ep},
\frac{t}{\ep^{5/2}}}\)
&=& \lim_{\ep \to 0}~\inv{\ep^{1/2}}\
\min_y\({ \Phi(y) +\ep^{5/2} \frac{(x\ep^{-1}-y)^2}{2t} }\) 
\non\\ 
&=&\min_y\({ \Phi_0^{**}(y) + \frac{(x-y)^2}{2t} }\) 
\label{wrong}
\fin
where $\Phi_0^{**}(x)\equiv \lim\limits_{\ep \to 0}
\inv{\ep^{1/2}}\ \Phi\({\frac{x}{\ep}}\)$ is the rescaled initial 
potential with the two point function $G^{**}(x) \equiv 
\lim\limits_{\ep \to 0}\inv{\ep}  G\({\frac{x}{\ep}}\)
= \bar D~ \de(x)$ with $\bar D=\int dx\, G(x)$. 
It has dimension $h=-\half$. Such a limit (\ref{inter}) would 
correspond to an intermediate regime with characteristics 
length $l^*(t)\sim t^{2/5}$ smaller than
Kida's length $l_K(t)\sim t^{1/2}$. The problem, however,
is that the realizations of the white noise $\Phi_0^{**}$ 
are distributional and the last expression in (\ref{wrong})
is ill-defined. Clearly a finer analysis is required to decipher 
cases in which a limit of the type (\ref{inter}) leads
to a self-similar asymptotic distribution with $h=-\frac{1}{2}$
as the one constructed above. In fact, we expect the latter 
to appear if the initial potentials have values uniformly 
bounded below, in analogy with the extreme statistics problem.
\vskip 0.2cm

Identical constructions, arguments and speculations could 
be done in higher dimensions. For example, in dimension $d$ 
the delta function has dimension $h=-\frac{d}{2}$, and
there exists a self-similar solution with this scaling dimension.
It corresponds to the characteristic length $l(t)\sim t^\al$ 
with $\al= \frac{2}{d+4}$.
\vskip 0.2cm

Finally, we feel that it could be worth-while to adapt 
the renormalization group techniques to analyze this type 
of large time behavior.

\end{document}